\documentclass[twocolumn,showpacs,preprintnumbers,amsmath,amssymb]{revtex4}
\usepackage{amsmath}
\usepackage{amsfonts}
\usepackage{amssymb}
\usepackage{graphicx}%
\usepackage{verbatim}   
\usepackage{color}      
\usepackage{subfigure}  
\usepackage{hyperref}   
\usepackage{dcolumn}
\usepackage{bm}
\begin{document}
\title{Euler-Lagrangian dynamics to the physical interpretation with granular constraints for MD simulations}

\author{Kyoung O. Lee}
\author{Robin P. Gardner}%
 \email{gardner@ncsu.edu}
\affiliation{Center for Engineering Applications of Radioisotopes\\ Department of Nuclear Engineering \\ North Carolina State University, Raleigh, NC 27695-7909}%

\begin{abstract}
 In this article, Euler-Lagrangian dynamics explain that the two particle interaction has non-conservative forces about the frame of the center of mass. This interpretation clarifies the underlying interaction and the system descriptions become advantages for MD simulations. 
\end{abstract}

\keywords{Molecular Dynamics, Simulation, DEM, Hertz, granular, Euler-Lagrangian}

\maketitle

A Molecular Dynamics (MD) Simulation is based on Newton's second law on time
evolution within the framework of classical mechanics. The motion of granular
dynamics can be described by Euler-Lagrangian dynamics. From a comprehensive
point of view, the Euler-Lagrangian dynamics is presented to clearly and
easily identify the pair interaction and contacting time among the individual
particles. The kinetic energy and the potential energy are used to obtain the
Lagrangian of the particle with the generalized coordinate and its associated
granular interaction term. In this work, the system of N particles was the
only constraint used to move the particles through the external gravitational
force and the energy dissipated in friction. 

For a granular system of N particles moving in three dimensional space with
constraints, D'Alembert's principle requires knowledge of the constraint
forces, (typically and static), kinetic frictional force and particle sliding.
What was needed was a description of the system that makes compromises when
dealing with the constraint relations. D'Alembert's principle provides such a
description for systems involving holonomic or nonholonomic constraints in
allowing virtual displacements, $\delta\mathbf{r}_{i}=\underset{a}{\sum}%
\frac{\partial\mathbf{r}_{i}}{\partial q_{a}}\delta q_{a}$ in contrast with a
real displacement $d\mathbf{r}_{i}$. There are two types of constraints. The
first one is holonomic constraints which can be solved for using kinematics;
here there is no work $(\mathbf{F}_{i}^{(c)}\cdot\delta\mathbf{r}_{i}=0)$
because virtual displacements $\delta\mathbf{r}_{i}$ are orthogonal to the
corresponding constraint forces $\mathbf{F}_{i}^{(c)}$. The second is
nonholonomic constraints which must be solved for using dynamics; here
consequently we have a work term $(\mathbf{F}_{i}^{(c)}\times\delta
\mathbf{r}_{i}\neq0)$.

As a consequence of these constraint relations, this problem is solved by
choosing Lagrange's equation. The Euler--Lagrange equations, $L\left(
q_{a},{\dot{q}}_{a}\right)  $ in the generalized coordinates, $q_{a}$ for a
discrete system is%

\begin{equation}
\frac{d}{dt}\frac{\partial L}{\partial{\dot{q}}_{a}}-\frac{\partial
L}{\partial q_{a}}=Q_{a}%
\end{equation}
Here $L\equiv T-V-U$ is the Lagrangian of the system.

The generalized constraint forces, $Q_{a}$ are derived from D'Alembert's
principle for the virtual work for applied forces. Two nonconservative
components are associated with respect to ${\dot{q}}_{a}$ and ${q}_{a}$. The
generalized constraint forces on the right-hand side are given by%

\begin{equation}
Q_{a}\equiv\mathbf{F}_{ij}^{(c)}\cdot\frac{\partial\mathbf{r}_{ij}}{\partial
q_{a}}=\mathbf{F}_{ij}^{(f)}\cdot\frac{\partial\mathbf{r}_{ij}}{\partial
q_{a}}-\frac{\partial\mathcal{F}}{\partial{\dot{q}}_{a}}%
\end{equation}
where $\mathbf{F}_{i}^{(f)}$ is only for the friction force and $\mathcal{F}$
is known as Rayleigh's dissipation function $\mathcal{F}=\frac{1}{2}%
\underset{a,b}{\sum}c_{ij}{\dot{q}}_{a}{\dot{q}}_{b}$ (if $c_{ij}=c_{ji}$ ,
the symmetric tensor).

The Lagrange equation with Rayleigh's dissipation function and contact
friction becomes%

\begin{equation}
\frac{d}{dt}\frac{\partial L}{\partial{\dot{q}}_{a}}-\frac{\partial
L}{\partial q_{a}}+\frac{\partial\mathcal{F}}{\partial{\dot{q}}_{a}}%
=Q_{a}^{(f)}%
\end{equation}
where $Q_{a}^{(f)}=\mathbf{F}_{ij}^{(f)}\cdot\frac{\partial\mathbf{r}_{ij}%
}{\partial q_{a}}$. The generalized constraint force is presented by a
frictional force.

In the case in which particles were assumed to be the spherical rigid bodies
located in the gravitational field, the generalized coordinates were decoupled
into the center of mass coordinates (the axes of the inertial coordinate
system) and the Euler angles $(\theta,\phi,\psi)$. Since the spherical rigid
body is symmetrical for the Euler angles, the moment of inertia for an ideal
homogeneous sphere of radius $R$ is $I=\frac{2}{5}mR^{2}$ is about the
principal axis of the center of mass. From the relations above we can compute
one term relating the kinetic energy $T$ of the translational motion and the
spin rotation itself about the center of mass. Another term relating the
potential energy of the gravity potential $V $ and a pairwise interaction
potential $U\left(  r_{ij}\right)  $, is given where the distance between the
centers of the two particles is $r_{ij}=\left\vert \mathbf{r}_{i}%
-\mathbf{r}_{j}\right\vert $ with respect to a fixed frame and $\mathbf{r}%
_{i}$, $\mathbf{r}_{j}$ is the relative position vector of the center of mass
$m_{i}$, $m_{j}$.

\begin{figure}
\centering
  \includegraphics[width=0.5\textwidth]{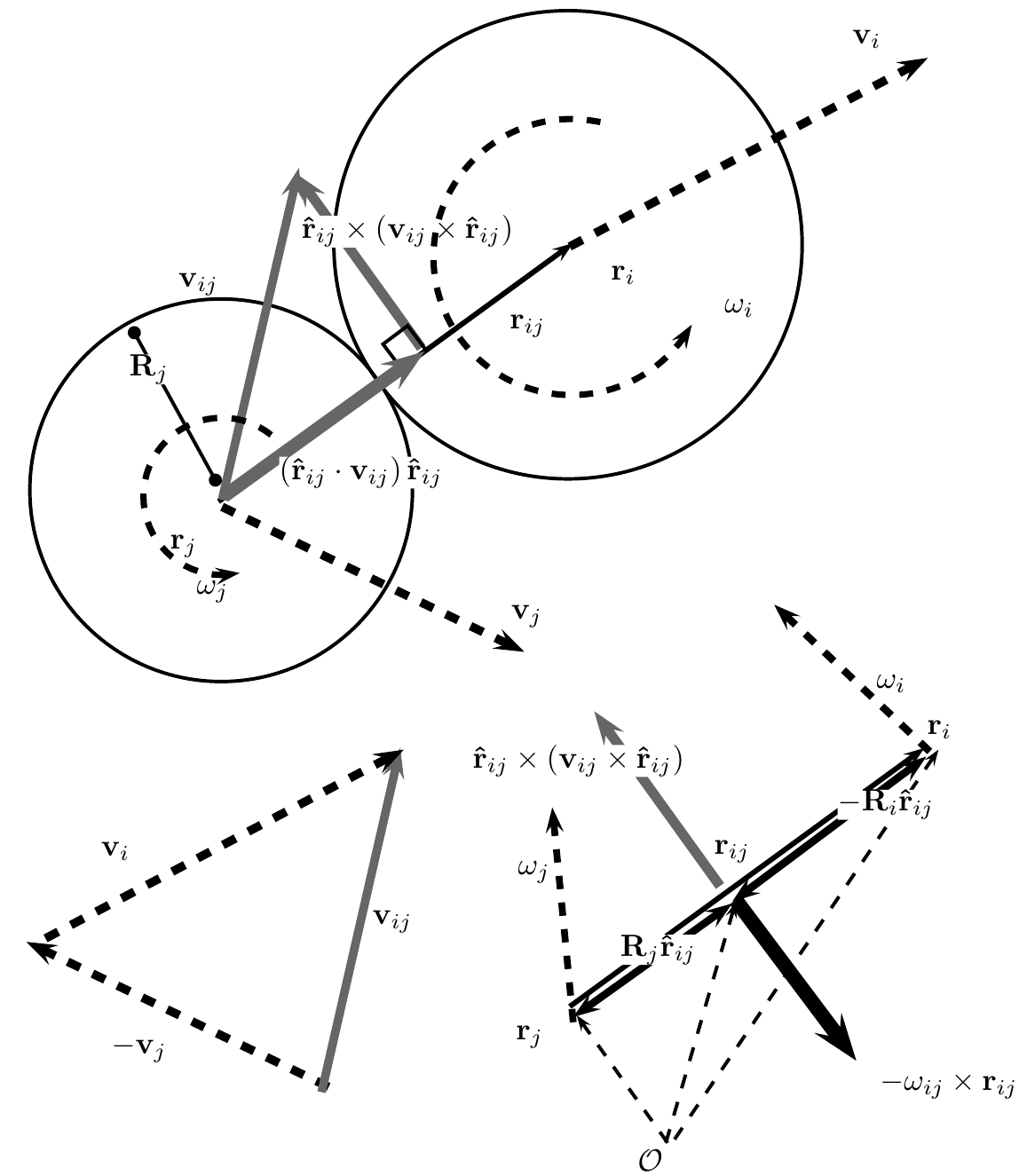}
\caption{Schematic of the vector analysis contacting on particle acting $j\rightarrow i$}
  \label{figure1}
\end{figure}

The distance between particles $i$ and $j$ about a fixed point $O$, the
relative position about the normal and tangential direction, is independent.
Above all, the relative angular velocity in the rotating frame is contributed
to the angular displacement from the equation%

\begin{equation}
\omega_{i}\times\frac{R_{i}}{R_{i}+R_{j}}\mathbf{r}_{ij}+\omega_{j}\times
\frac{R_{j}}{R_{i}+R_{j}}\mathbf{r}_{ij}=\omega_{ij}\times\mathbf{r}_{ij}%
\end{equation}

The relationship of the overlap displacement is defined as
\begin{equation}
\delta{r}_{ij}{=}\left(  \mathbf{\hat{r}}_{ij}\cdot\delta_{ij}\right)
\mathbf{\hat{r}}_{ij}+\mathbf{\hat{r}}_{ij}\times\left(  \delta_{ij}%
\times\mathbf{\hat{r}}_{ij}\right)  -\int\mathbf{\omega}_{ij}\times
\mathbf{\hat{r}}_{ij}d\tau
\end{equation}
Here the center of angular velocity is $\omega_{ij}\equiv\left(  R_{i}%
\omega_{i}+R_{j}\omega_{j}\right)  /\left(  R_{i}+R_{j}\right) $, with angular
velocities $\omega_{i}$ and $\omega_{j}$ in a rotating coordinate frame with
each particle at the center of mass of the system during contact time $t_{c}$.

The relative normal displacement is
\begin{equation}
\delta{r}_{ij\parallel}=\left(  \mathbf{\hat{r}}_{ij}\cdot\delta_{ij}\right)
\mathbf{\hat{r}}_{ij}%
\end{equation}
and the relative tangential displacement is
\begin{equation}
\delta{r}_{ij\perp}=\mathbf{\hat{r}}_{ij}\times\left(  \delta_{ij}%
\times\mathbf{\hat{r}}_{ij}\right)  -\int_{t_{c}}\mathbf{\omega}_{ij}%
\times\mathbf{\hat{r}}_{ij}d\tau
\end{equation}

A comparison with the above gives the relative velocity as%

\begin{align*}
\left.  \frac{d\mathbf{r}_{ij}}{dt}\right\vert _{s}  &  =\mathbf{\dot{r}}%
_{ij}-\mathbf{\omega}_{ij}\times\mathbf{r}_{ij}\\
&  =\left(  \mathbf{\hat{r}}_{ij}\cdot\mathbf{v}_{ij}\right)  \mathbf{\hat{r}%
}_{ij}+\mathbf{\hat{r}}_{ij}\times\left(  \mathbf{v}_{ij}\times\mathbf{\hat
{r}}_{ij}\right)  -\mathbf{\omega}_{ij}\times\mathbf{r}_{ij}%
\end{align*}
where ${\hat{r}}_{ij}=\mathbf{r}_{ij}/r_{ij}$ and $\mathbf{r}_{ij}%
=\mathbf{r}_{i}-\mathbf{r}_{j}$, the relative velocity in term of $j$.

The relative velocity $\mathbf{\dot{r}}_{ij}=\mathbf{v}_{ij}=\mathbf{v}%
_{i}-\mathbf{v}_{j}$ has the components of the relative normal displacement
\begin{equation}
\mathbf{v}_{ij\parallel}=\left(  \mathbf{\hat{r}}_{ij}\cdot\mathbf{v}%
_{ij}\right)  \mathbf{\hat{r}}_{ij}%
\end{equation}
and the relative tangential velocity is given by
\begin{equation}
\mathbf{v}_{ij\perp}=\mathbf{\hat{r}}_{ij}\times\left(  \mathbf{v}_{ij}%
\times\mathbf{\hat{r}}_{ij}\right)  -\mathbf{\omega}_{ij}\times\mathbf{r}_{ij}%
\end{equation}
where $\mathbf{\hat{r}}_{ij}\times\left(  \mathbf{v}_{ij}\times\mathbf{\hat
{r}}_{ij}\right)  =\mathbf{v}_{ij}-\left(  \mathbf{\hat{r}}_{ij}%
\cdot\mathbf{v}_{ij}\right)  \mathbf{\hat{r}}_{ij}$ .

When granular particles have pairwise collisions, they collide instantly and
have an infinitesimally small time for the Hertz's potential. This impulse
imparts interactions between each particle. Figure shows analytic geometry,
before instantaneous deformation, at which point the bodies touch at an
adjacent point. The curvatures of the two bodies are represented with radii of
curvature $R_{i}$ and $R_{j}$. At the contact area, the bodies are compressed
with smooth curvature. The shape of the particle is changed by the curvature
due to the elastic restoration force. Previously, Hertz's theory had been used
with a quadratic equation. However, contact pressure provides displacement
under the surface. This contact pressure is distributed throughout the
curvature. The pressure distribution function acts at each contact radius $a$.
The distribution of normal stress in the contact area as a function of
distance was also reported \cite{2}. The granular particles behave like
hard sphere which contact one another with a deformable contact force. The
forces between the two particles can be approximated well by Hertz theory of
elasticity. The granular interaction potential is a power law from Hertz
theory. Hertz contact is presented to determine each granular surface
interaction for granular flows. The generalized expression for the normal
contact force acting on a granular body, in which the governing equation is
either loaded in tension or in compression, is obtained from the stress and
strain relation of the material. The Hertz potential is derived from the
results of stress-strain analysis that are attributed to granular flow
sensitivity. Granular particle interactions have a virtual depth in the
relation with a remarkable normal contact of relative deformation in terms of
a tangential contact by a friction on a contact surface. A substantial
solution to the problem of a normal contact has been known as Hertz's theory.

\begin{figure*}
\centering
  \includegraphics[width=1.0\textwidth]{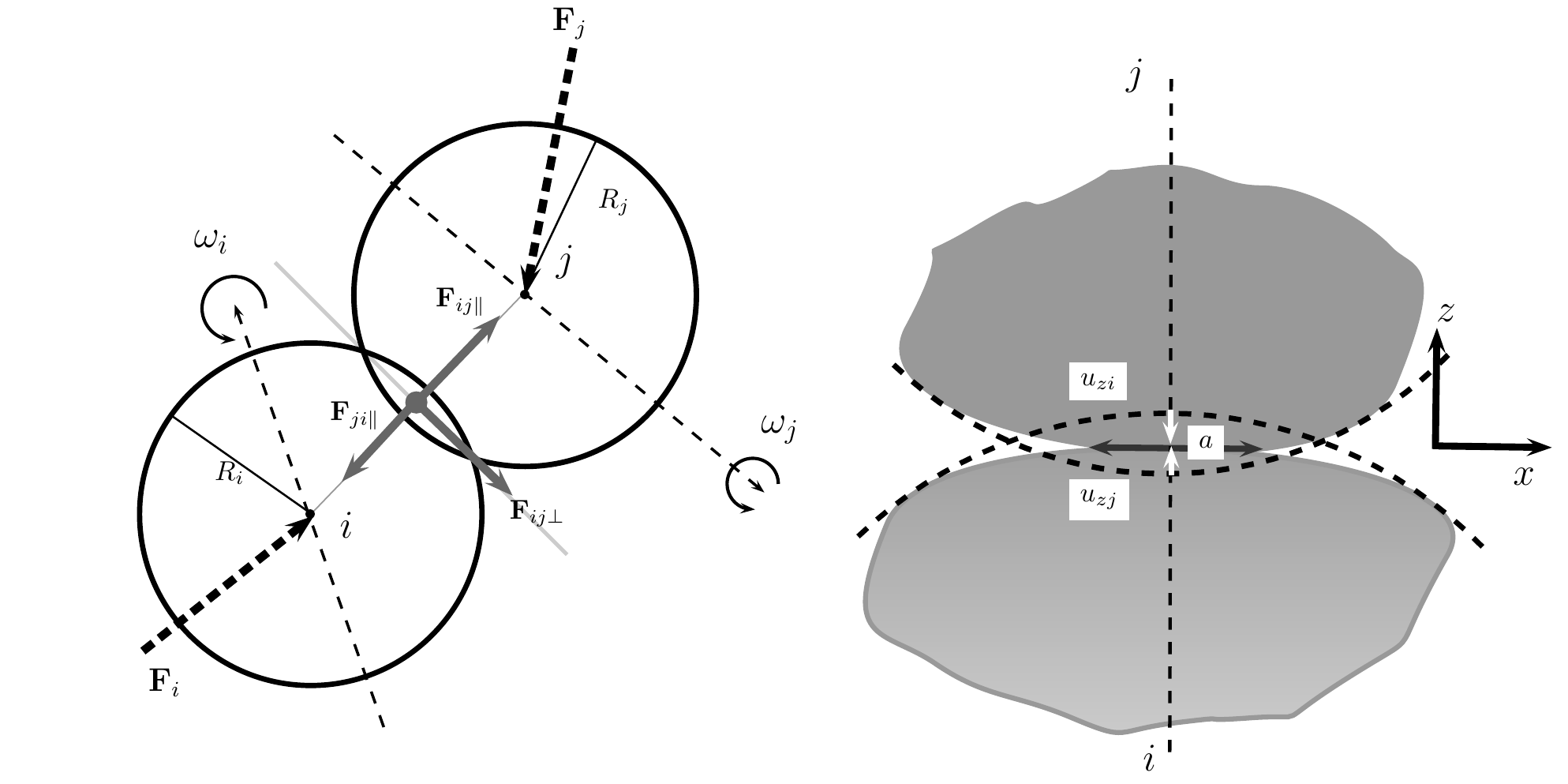}
\caption{Schematic of the interaction between two particles at
contact with position vectors $\mathbf{r}_{i}$, $\mathbf{r}_{j}$: The left figure of a set of a overlap shape and the right figure of the deformable shape of Hertz-Mindlin force}
  \label{figure2}
\end{figure*}

We are led to the relations of the overlap displacement as defined by
$\delta{r}_{ij}$ where the normal displacement $\delta r_{ij\parallel}\equiv
R_{i}+R_{j}-\left\vert \mathbf{r}_{i}-\mathbf{r}_{j}\right\vert $ is the depth
of indentation or normal overlap for the contact between the two bodies of
radii $R_{1}$ and $R_{2}$. When $\delta r_{ij\parallel}>0$, a contact force is
generated that otherwise would have zero potential. The noncohesive granular
interaction is described by Hertz-Mindlin potential between particles,%

\begin{equation}
U\left(  r_{ij}\right)  =\frac{2}{5}k\delta r_{ij}^{5/2}%
\end{equation}
where $k_{n}=\frac{4}{3}E^{\ast}\sqrt{R^{\ast}}$ is the stiffness of the
pairwise interaction proposed in Hertz theory (Hertz 1895) and the tangential
term $k_{t}=8G^{\ast}\sqrt{R^{\ast}}$ is the tangential force (Mindlin 1949)
with the effective Radius $R^{\ast}=R_{i}R_{j}/\left(  R_{i}+R_{j}\right)  $
and the effective Young's modulus $E^{\ast}=\left(  \frac{1-\nu_{i}^{2}}%
{E_{i}}+\frac{1-\nu_{j}^{2}}{E_{j}}\right)  ^{-1}$, $E$ the Young's modulus
and the effective Shear modulus is $G^{\ast}=\left(  \frac{2\left(  2-\nu
_{i}\right)  \left(  1+\nu_{i}\right)  }{E_{i}}+\frac{2\left(  2-\nu
_{j}\right)  \left(  1+\nu_{j}\right)  }{E_{j}}\right)  ^{-1}$. Poisson's
ratio is $\nu$ for the contacting particle $i$ and $j$ respectively
\cite{1,3,4}.

Comparisons of this with the Lagrangian obtained previously can be expressed
by the gravitational potential%

\begin{equation}
V\left(  r_{i}\right)  =m_{i}\mathbf{r}_{i}\cdot\mathbf{g}%
\end{equation}
where $\mathbf{g}$ is the gravitational acceleration vector.

The pair dissipative function can be derived from Rayleigh's dissipation function%

\begin{equation}
\mathcal{F}=\frac{1}{2}\gamma m^{\ast}\delta\dot{r}_{ij}^{2}%
\end{equation}
where $\gamma$ is a dissipative constant or viscous damping coefficient given
a coefficient of normal restitution $e$ for a contacting time with normal and
tangential contact.

When the particle $i$ is moving to another particle $j$ at contacting time, it
follows that the force of constraint exerted by kinetic friction ($\mu_{k}$,
the kinetic coefficient of friction) of each particle has done real work after
the collision. Thus the virtual displacement, $\delta\mathbf{r}$, is tangent
to the other particle and orthogonal to the constraint force. If both
particles are at a rest state $\left(  \mu_{s}\leq\mu_{k}\right)  $, the
sphere at contact experiences the only force on the pebble from the static
friction ($\mu_{s}$, the static coefficient of friction) which is
perpendicular to the curved part of particle. The normal and tangential
direction is independent, and then each term must be separate. The motion of
the $i$th particle is given by a contact force and a damping force during
collisions. If the tangential force is generated, the generalized constraint
force is $Q_{a}^{(f)}=-sign\left(  \delta r_{ij\perp}\right)  \min\left(
\mu\left\vert \mathbf{F}_{ij\parallel}\right\vert ,\left\vert \mathbf{v}%
_{ij\perp}\right\vert \right)  \mathbf{\hat{r}}_{ij\perp}$.

Thus, the Lagrange's equation of the net binary system can be expressed in
term of $r_{ij}$%

\begin{equation}
L=\frac{1}{2}m^{\ast}\left\vert \delta{\dot{r}}_{ij}\right\vert ^{2}-U\left(
r_{ij}\right)
\end{equation}
where $m^{\ast}=m_{i}m_{j}/\left(  m_{i}+m_{j}\right)  $ is the reduced mass
of the system.

As a consequence of the decoupling, the Lagrange equations of motion separate
into two equations, one for the center of mass coordinates and another for the
Euler angles. Each set can be analyzed independently to apply this
prescription to a two body problem for a virtual displacement. The present
simulations use a normal and tangential contact model comprised of the following:%

\begin{equation}
\left\vert \delta{\dot{r}}_{ij}\right\vert ^{2}=\left\vert \delta{\dot{r}%
}_{ij\parallel}\right\vert ^{2}+\left(  1+\frac{1}{\zeta}\right)  \left\vert
\delta{\dot{r}}_{ij\perp}\right\vert ^{2}%
\end{equation}
where $\zeta=\left(  \frac{1}{m_{i}\tilde{I}_{i}}+\frac{1}{m_{j}\tilde{I}_{j}%
}\right)  ^{-1}$ from the particle moment of inertia $\tilde{I}_{i}%
=I_{i}/m_{i}R_{i}^{2}$ and $\tilde{I}_{j}=I_{j}/m_{j}R_{j}^{2}$ about the
torque acting on two particle.

Finally, we can implement Euler-Lagrangian dynamics to find out the governing
equation and contact time used to simulate the MD method. Although the
theoretical approach is Hertz's contact to the Lagrangian solution, problems
arise including general friction through contacting area where $\mathbf{F}%
_{ji}=-\mathbf{F}_{ji}$, Newton 3rd law and the tangential component
$\mathbf{F}_{ij\perp}$ and $\delta{r}_{ij\perp}$ which can be described to
meet the requirement of Coulomb yield criterion if $\left\vert \mathbf{F}%
_{ij\perp}\right\vert <\mu_{s}\mathbf{F}_{ij\parallel}$. Until the particles
are separated, the potential calculation is performed for the interaction only
once on each pair of particles. The equations of motion follow directly from
the Lagrangian formulation described above.

The normal component of the contact force can be written as%

\begin{equation}
\mathbf{F}_{ij\parallel}{=}{k}_{n}\delta{r}_{ij\parallel}^{3/2}-\gamma
_{n}m^{\ast}\mathbf{v}_{ij\parallel}%
\end{equation}

The shear component of the contact force can be written as%

\begin{equation}
\mathbf{F}_{ij\perp}=-{k}_{t}\sqrt{\delta{r}_{ij}}\delta{r}_{ij\perp}%
-\gamma_{t}m^{\ast}\mathbf{v}_{ij\perp}%
\end{equation}

The equation of motion will be expressed in normal and tangential directions
on contact condition, $\left\vert \mathbf{F}_{ij\perp}\right\vert >\mu
_{s}\mathbf{F}_{ij\parallel}$, $\mathbf{F}_{ij\perp}=-\mu_{k}\mathbf{F}%
_{ij\parallel}\mathbf{\hat{v}}_{ij\perp}$ and $\gamma_{t}m^{\ast}%
\mathbf{v}_{ij\perp}$ which generates the rolling friction from the center of
particle $j$ to the center of particle $i$. This component in the direction of
motion is attributed by performance of the Hertz-Mindlin theory to improve the
terms of the tangential force. The simulation will demonstrate the essential
results of the contact deformable interaction. Two body forces are represented
by the potential, which is enough to permit the surface effects. The governing
equation, constructed from two-body functions, is applied to the many body
system problem with a contact dynamic using a MD simulation (or DEM).

\end{document}